# The Noise of the Charge Density Waves in NbSe$_3$ Nanowires – Contributions of Electrons and Quantum Condensate

Subhajit Ghosh[1], Sergey Rumyantsev[2], and Alexander A. Balandin[1,3*]

[1]Department of Materials Science and Engineering, University of California, Los Angeles, California 90095 USA

[2]CENTERA Laboratories, Institute of High-Pressure Physics, Polish Academy of Sciences, Warsaw 01-142, Poland

[3]California NanoSystems Institute, University of California, Los Angeles, California 90095 USA

* Corresponding author (A.A.B.): balandin@seas.ucla.edu ; web-site: https://balandin-group.ucla.edu/

## Abstract

Low-frequency electronic noise in charge-density-wave van der Waals materials has been an important characteristic, providing information about the material quality, phase transitions, and collective current transport. However, the noise sources and mechanisms have not been completely understood, particularly for the materials with a non-fully gapped Fermi surface where the electrical current includes components from individual electrons and the sliding charge-density wave. We investigated noise in nanowires of quasi-one-dimensional $NbSe_3$, focusing on a temperature range near the Pearls transition $T_{P1}$ ~145 K. The data analysis allowed us to separate the noise produced by the individual conduction electrons and the quantum condensate of the charge density waves before and after the onset of sliding. The noise as a function of temperature and electric bias reveals several intriguing peaks. We explained the observed features by the depinning threshold field, the creep and sliding of the charge density waves, and the possible existence of the hidden phases. It was found that the charge density wave condensate is particularly noisy at the moment of depinning. The noise of the collective current reduces with the increasing bias voltage in contrast to the noise of the individual electrons. Our results shed light on the behavior of the charge density wave quantum condensate and demonstrate the potential of noise spectroscopy for investigating the properties of low-dimensional quantum materials.

I. **Introduction**

Low-dimensional layered materials, which include quasi-one-dimensional (1D) and quasi-two-dimensional (2D) van der Waals materials, are attracting growing attention owing to their intriguing physics and unique functionalities for future practical applications[1–7]. Particularly interesting are quasi-1D and quasi-2D materials that reveal charge-density-wave (CDW) phases at various temperatures; in some cases, above-room temperature (RT)[8–17]. The additional impetus to study such low-dimensional CDW material systems comes from the realization that in certain strongly anisotropic van der Waals materials, the CDW phenomena coexist with the topological phases[18–21]. The CDW ground state is a condensate of electrons that differ in momentum by $2k_F$, or, in the equivalent interpretation, it is a condensate of $2k_F$ phonons ($k_F$ is the Fermi wave vector). This is similar to a superconductor, which is a condensate of electron Cooper pairs of opposite spin and momentum. A collective transport mode of the quantum condensate, in principle, can allow CDW to slide relative to the lattice and carry charge without friction *via* zero-resistance Frohlich current[9,11,22–24]. However, in real crystals, the CDW quantum condensate is always pinned to impurities, defects, and dislocations, and a zero-resistance current has never been observed experimentally. To initiate the CDW depinning and sliding, a finite electric field, $E_T$, has to be applied[11,25,26].

Electrical current fluctuations, or noise, have always been an essential component of CDW research. Noise contains information about the material quality, defects, impurities, phase transitions, charge carrier recombination, and charge transport. Using terminology, conventional in the CDW field, one distinguishes the narrow-band noise (NBN) and broad-band noise (BBN)[27–31]. It is understood that NBN is a signature of the CDW sliding. It is the AC component appearing in the output signal from the CDW material under DC bias owing to the oscillating collective current of CDW[27,30,32,33]. The term BBN, specifically in the context of CDW materials, implies the low-frequency noise (LFN) or excess noise observed in all metals and semiconductors[28,34–36]. Despite decades of investigations, the noise sources and their mechanisms in CDW materials have not been completely understood, particularly for the materials with a non-fully gapped Fermi surface where the electrical current includes components from individual electrons and the current of the sliding CDWs. The revival of the CDW research field and technological improvements call for the in-depth examination of the noise characteristics of CDW materials to gain a deeper understanding of their origins and mechanisms. This is important for monitoring the evolution of the CDW quantum condensate phases and for the practical realization of proposed applications of CDW materials in future low-power and radiation-hard electronics[13,37–39]. Noise spectroscopy has also proven to be a valuable approach for fundamental studies of quantum materials, and its application to CDW materials may bring a lot of physics insights[34,35,40].

For this investigation, we selected niobium triselenide ($NbSe_3$) – a quintessential CDW material with a quasi-1D crystal structure. It has been studied extensively in the form of bulk crystals or large-diameter whiskers[9,11,41,42]. This material undergoes two CDW transitions, one at $T_{P1}$ ~145 K and one at $T_{P2}$ ~59 K, which are attributed to the developing energy gaps in different sections of the Fermi surface. The specific of $NbSe_3$ is that certain parts of its Fermi surface always remain un-gapped, so the current comprised of individual electrons, *i.e.,* normal carrier conduction, is always present in addition to the collective conduction current owing to the CDW sliding. At the applied electric field, $E$, which is below the threshold field, $E_T$, the electric current consists of individual electrons only, but the formation of CDW below the first Peierls CDW transition temperature, $T_{P1}$, affects the conduction. At the electric field $E>E_T$ and $T<T_{P1}$, the total current, $I_T$, is a sum of the collective CDW current, $I_{CDW}$, and the individual electrons current, $I_S$, *i.e.,* $I_T=I_{CDW}+I_S$. This mixture of the current due to individual electrons and sliding of the CDW quantum condensate makes the noise characteristics of $NbSe_3$ particularly intriguing.

Prior studies of electronic noise in $NbSe_3$, which used bulk crystal samples rather than nanowires, addressed both BBN and NBN[31,43–47]. It was found that BBN noise had the spectral density, $S\sim1/f^{\gamma}$, with the power factor $\gamma$ deviating strongly from unity, in the range $0.4\leq\gamma\leq0.8$[45]. Another study of noise over a broad temperature range reported that at temperatures above $T_{P2}$, the noise followed the $1/f^{\gamma}$ trend with $\gamma$~0.8, while at low temperatures, it revealed $\gamma$~1.8[43]. In conventional metals and semiconductors, LFN typically has a power factor $\gamma$ equal to or close to unity[48]. In semiconductors, the deviation from the $1/f$ dependence was conventionally attributed to the appearance of the Lorentzian bulges of the generation–recombination (G-R) noise superimposed on the $1/f$ background[49]. While there was an agreement in published reports on the strong deviation of power factor $\gamma$ from unity in $NbSe_3$, the interpretations of it differ substantially.

There was also a lack of understanding of the difference between the noise produced before and after the CDW de-pining.

For this study, we used exfoliated nanowires of NbSe$_3$ as the channels in the device test structures. The amplitude coherence length in CDW materials is relatively small, on the order of ~10 nm, while the phase coherence length is large and can be up to ~10 μm[42,50,51]. The phase coherence length in the cross-plane direction is ~1 μm[52,53]. Using nanowires with cross-sectional dimensions on the order of the amplitude coherence length and length on the order of the phase coherence length may allow one to elucidate better the effects due to the CDW formation and sliding. In the present work, we ask the questions: How noisy are the CDW quantum condensate phases when they are de-pinning *vs* when their sliding is well established in the incommensurate CDW (IC-CDW) phase? Does the collective current of the sliding quantum condensate produce more or less noise than the current of individual electrons? We observed intriguing features in the noise spectral density as a function of temperature and applied electric field. Our data analysis allowed us to separate the noise produced by the individual conduction electrons and the sliding of the CDW quantum condensate and offer a hypothesis for the peaks in the noise spectra as functions of temperature and electric field. We believe that the unusual features found in the noise spectra can stimulate theoretical developments in the field of quantum materials and their applications.

## II. Test Structures and Current-Voltage Characteristics

We prepared nanowires of NbSe$_3$ by mechanically exfoliating small pieces of high-quality bulk NbSe$_3$ samples (2D Semiconductors) on top of clean Si/SiO$_2$ substrates ((University Wafer, p-type Si/SiO$_2$, <100>). The quality of the bulk samples was assessed using Raman spectroscopy (Renishaw inVia). The room-temperature Raman measurements were conducted using λ= 488 nm laser excitation with a laser power of less than ~ 1 mW to avoid damage to the sample (see Figure 1 (a)). The observed Raman peaks are in good agreement with the literature, confirming the crystal structure and purity of the material[54]. For device fabrication, several multi-contact test structures, also referred to as devices, were fabricated using the standard fabrication techniques[40]. We started by cleaning the small pieces of Si/SiO$_2$ substrates several times using acetone and isopropyl alcohol (IPA) and rinsing them with deionized (DI) water. A conventional mechanical exfoliation (Nitto tape) was employed to exfoliate the bulk materials into nanowire structures on top of the clean substrates. Figure 1 (b) shows the scanning electron microscopy (SEM) image of the exfoliated nanowires with rectangular cross-sections. Additional microscopy images are provided in Supplemental Figures S1 and S2.

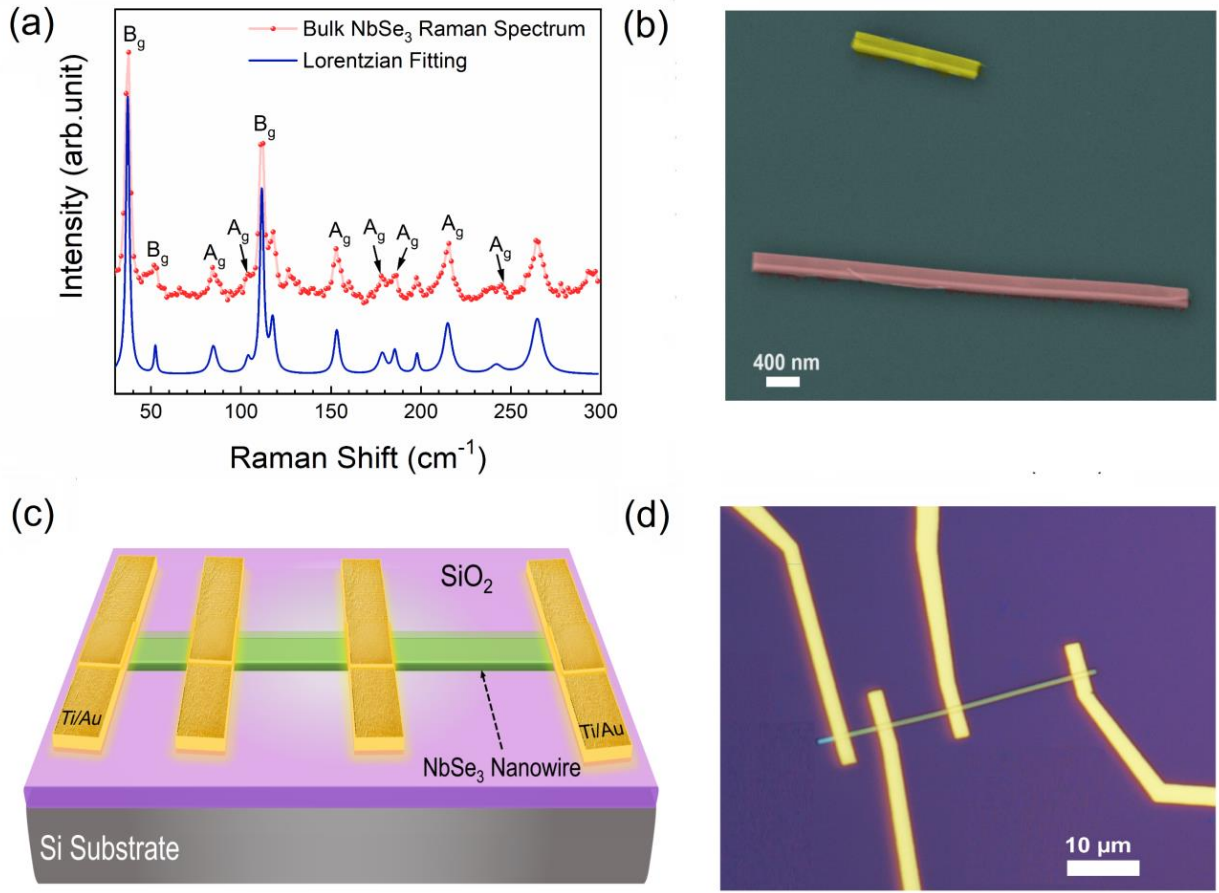

**Figure 1:** (a) Raman spectrum of a bulk $NbSe_3$ sample measured at room temperature using $\lambda=$ 488 nm laser excitation. (b) SEM image of exfoliated $NbSe_3$ nanowire. Pseudo-colors are used for clarity. The exfoliated nanowires typically have rectangular cross-sections, as confirmed by SEM. (c) Schematic diagram of the $NbSe_3$ nanowire test structure on $Si/SiO_2$ substrate. (d) Optical microscopy image of a representative multi-channel $NbSe_3$ nanowire test structure.

For metal contact fabrication across the nanowire lengths, the substrates containing nanowires were first spin-coated using PMMA (Kayaku Advanced Materials, 495 PMMA A6)[39,40,55]. Next, the electron-beam (e-beam) lithography (EBL) technique was used to write patterns on the coated

PMMA layer for contact electrodes and pads. After the e-beam exposure, the substrates were submerged into a developer solution (Microposit 351) and IPA for 20 sec each. The development technique ensured the removal of the PMMA layer exposed by e-beams. The development process was followed by a plasma-based cleaning process for 30 sec to remove any chemical residue in the exposed patterns. The metal edge contacts were created by depositing a combination of Ti/Au (20 nm/150 nm) metals using an e-beam evaporation (EBE) system. In the next step, the substrates were submerged into an acetone solution for the lift-off process. Finally, $NbSe_3$ nanowire devices with varying channel lengths between 2 μm - 30 μm were fabricated for electrical testing. The schematic image of a $NbSe_3$ multi-contact nanowire test structure is shown in Figure 1 (c). An optical microscopy image of a representative device is provided in Figure 1 (d).

The current-voltage (*I-V*) characteristics were measured in a conventional two-terminal configuration inside a cryogenic probe station (Lakeshore TTPX) in the temperature range from 77 K to 300 K. The measurements were conducted using a semiconductor parameter analyzer (Agilent B1500A). Figure 2 (a) shows the *I-V* characteristics of the $NbSe_3$ nanowire device at room temperature (RT). The *I-V* behavior at RT is Ohmic, as evidenced by the linear current response as a function of voltage. The linear *I-V* confirms the metallic nature of the material at RT. It also indicates the high quality of contacts between the channel and the metal electrodes, which is essential for reliable noise studies. The *I-V* characteristics change as the temperature goes below the $T_{P1}$~145 K, which is the first CDW phase transition temperature of $NbSe_3$[56–58]. One can see in Figure 2 (b) that the *I-V* characteristics have linear behavior at the lower bias region but become super-linear after the bias voltage exceeds the threshold for the CDW quantum condensate depinning. The observed nonlinear trend agrees with prior studies of bulk $NbSe_3$ crystals and is attributed to CDW depinning and sliding[47,59–61]. At T=120 K, the quasi-1D $NbSe_3$ material is in the IC-CDW phase, and after depinning the total measured current, $I_T$, consists of normal single charge carrier contribution, $I_S$, and the collective current of the sliding CDW condensate, $I_{CDW}$, so that $I_T=I_{CDW}+I_S$. Extrapolating the linear part of the *I-V* characteristics, one can plot the contributions of both current components separately (see Figure 2 (b)).

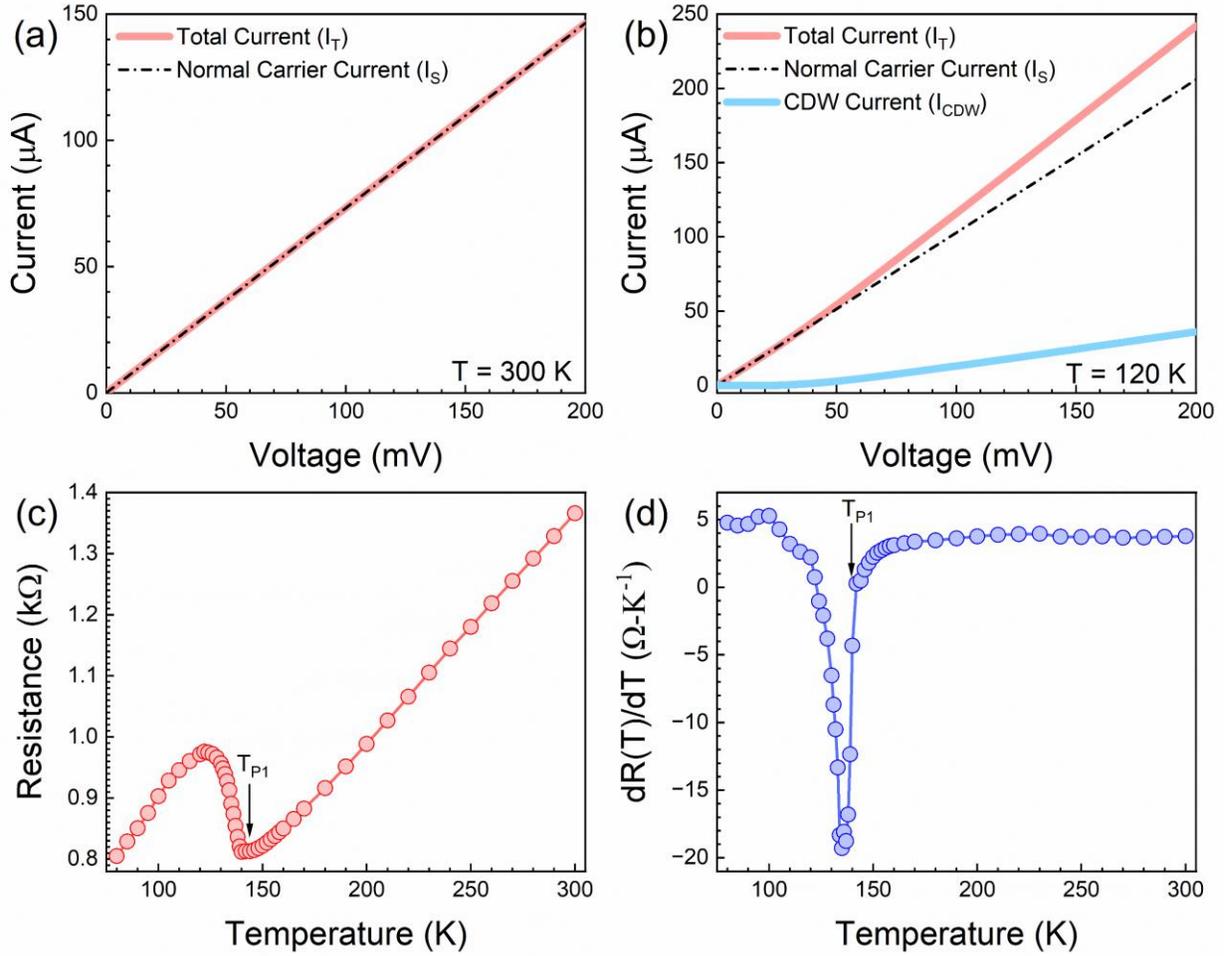

**Figure 2:** (a) Current-voltage (*I-V*) characteristics of the NbSe$_3$ nanowire device at room temperature (*T*=300 K). (b) Non-linear *I-V* characteristics of the same device at 120 K in the incommensurate CDW phase. The onset of non-linearity corresponds to the depinning and sliding of the CDW quantum condensate. (c) Resistance of the NbSe$_3$ device as a function of temperature measured in the heating cycle. The device shows a pronounced resistive anomaly near the CDW phase transition temperature $T_{P1} \approx 145$K. (d) The derivative characteristic, *dR/dT*, *vs.* temperature. The temperature where *dR/dT* = 0 corresponds to the onset of the CDW phase transition.

The resistance of the NbSe$_3$ channel, measured at low bias voltage, has a specific and well-understood dependence on temperature, as shown in Figure 2 (c) (see also Supplemental Figure S3 for data obtained for another device). Above $T_{P1}$~145 K, the material shows a typical metallic behavior, *i.e.,* the resistance increases with increasing temperature due to increasing electron–phonon scattering. At temperature $T_{P1}$~145 K, the material undergoes its first CDW phase transition with the formation IC-CDW phase of the quantum condensate, which is accompanied by a partial band gap opening along certain crystallographic directions[62]. At this point, a fraction of individual electrons condenses into a pinned CDW collective state, thus removing the charge carriers from the conduction band. As a result, the resistivity increases as the temperature decreases below ~145 K. The further decrease in temperature below 120 K brings a decrease in the resistivity due to an increase in the mobility of the remaining charge carriers. This is expected for metallic conductors where the phonon scattering limits the mobility of charge carriers. In Figure 2 (d), the first-order derivative of resistance with temperature, *dR/dT*, is plotted as a function of device temperature. The *dR/dT vs. T* plot reveals the transition temperature $T_{P1}$~145 K defined by the point where *dR/dT* = 0. The *I-V* measurements for a set of NbSe$_3$ nanowires confirm that we have test structures that reveal typical CDW characteristics of NbSe$_3$, in agreement with prior reports[47,59].

### III. Temperature Dependence of Noise in NbSe$_3$ Nanowires

The LFN measurements were carried out using a home-built noise measurement system. The measurement system comprises a 12 V DC battery connected in series with the device under test (DUT) and a load resistor. A potentiometer (POT) controls the voltage drop across the DUT and the load resistor. During the noise measurements, the output voltage fluctuations are transferred to a preamplifier (SR-560), which sends the amplified signal to a signal analyzer (Photon+). The signal analyzer converts the time domain voltage fluctuations to its equivalent noise power spectral density, $S_V$, through the Fourier transform. The obtained voltage spectral density is then converted to the equivalent short-circuit current spectral density, $S_I$. Our prior publications can reference our noise measurement system details[40,55,63–66]. Figure 3 (a) presents the current-referred noise spectral density, $S_I$, as a function of frequency, *f*, of NbSe$_3$ nanowire device for different current levels, measured at RT. The noise behavior at RT shows *1/f* dependence at all currents, confirming that the electronic noise in the material is of *1/f* flicker type without the Lorentzian component in the measured frequency range. The *1/f* noise behavior is expected for metals and most semiconductors[49,67]. Figure 3 (b) shows the current dependence of the noise spectral density, $S_I$, at a frequency *f* = 10 Hz for the given device at *T* = 300K. One can see in Figure 3 (b) that the $S_I$ *versus I* dependence reveals a quadratic dependence, *i.e.*, $S_I$~$I^2$, which is characteristic of linear resistors.

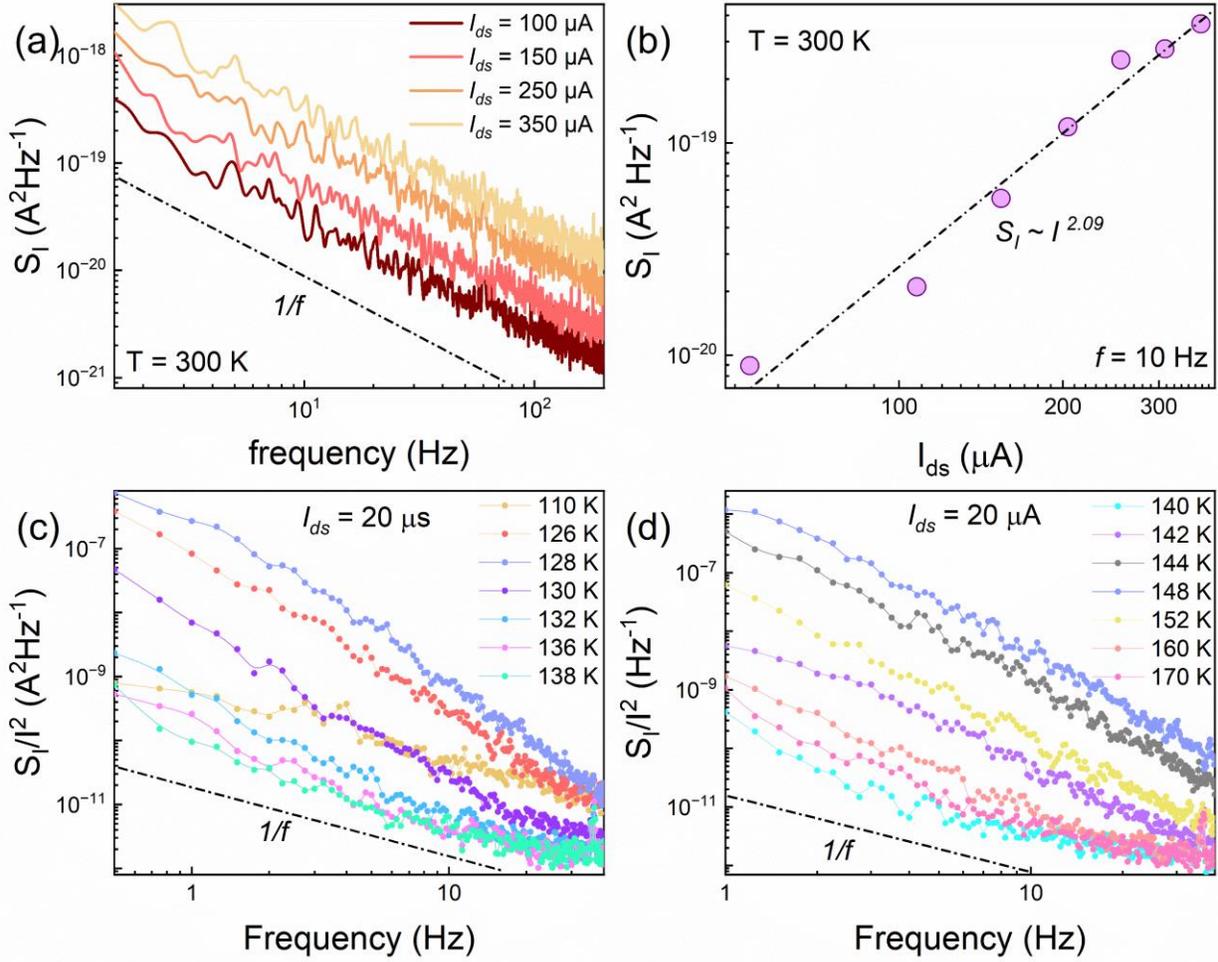

**Figure 3:** (a) Current noise spectral density, $S_I$ vs. frequency at $T$=300 K. The data are shown for different current levels in the NbSe$_3$ nanowire device. The spectral density shows *1/f* dependence at all the measured currents. (b) Current spectral density, $S_I$, at $f$ = 10 Hz and T = 300K plotted as a function of the device current. The $S_I \sim I^2$ dependence is expected for a linear resistor. (c) The normalized noise spectral density, $S_I/I^2$ vs. frequency measured at the fixed device current of $I_{ds}$ = 20 µA and temperatures in the range from 110 K to 138 K. (d) The same as in panel (c) but for a different temperature range from 140 K to 170 K. Note the evolution of the noise spectrum shapes as a function of temperature.

The systematic temperature-dependent noise measurements were conducted with several NbSe$_3$ nanowire devices at different current levels. Figures 3 (c) and 3 (d) show the normalized current noise spectral density, $S_I/I^2$, vs. frequency in a representative device at different temperatures and the fixed current $I_{ds}$ = 20 µA. At this low current, corresponding to the low bias voltage, the device *I-V* characteristics are in the linear region before the CDW depinning, as discussed previously. In Figure 3 (c), the noise data are presented for the temperature from 110 K to 138 K. The figure shows that the noise behavior is generally of $1/f$ type at lower temperatures but changes the shape in the intermediate temperature region, from $T=126$ K to $T=130$ K. The steeper spectra shapes are an indication of the Lorentzian type of the spectra with the characteristic corner frequency located outside the lowest measurable frequency. In Figure 3 (d), the noise data are presented for the temperature from 140 K to 300 K. The noise reveals a similar trend as in Figure 3 (c), with the Lorentzian tails at the intermediate temperatures and $1/f$ type dependence overall. The data presented in Figures 3 (c) and (d) suggest that the noise experiences multiple peaks as a function of temperature near the $T_{P1}$ temperature, associated with the appearance of the Lorentzian bulges. Supplemental Figure S4 shows similar noise behavior at other current levels.

To analyze the noise behavior near the CDW transition, we plotted the resistance and the normalized noise spectral density, $S_I/I^2$, as a function of temperature for the fixed values of current and frequency. In Figure 4 (a-d), the $S_I/I^2$ vs. $T$ dependence was plotted at four current levels and compared with the corresponding resistance $T$ dependence. Figure 4 (a) shows the $S_I/I^2$ vs. $T$ dependence at $I_{ds} = 20$ µA. The overall low noise level lies in the range from $10^{-10}$ Hz$^{-1}$ to $10^{-12}$ Hz$^{-1}$ except for the two intermediate regions where the noise shows sharp, well-distinguished peaks, with the noise values increasing by several orders of magnitude. The temperatures at which the peaks were observed revealed Lorentzian features in the noise spectral responses, as shown in Figure 3 (c-d) and the supplemental Figure S4. We observed consistent noise behavior at the other current levels of $I_{ds} = 30$ µA, 50 µA, and 100 µA, as shown in Figures 4 (b-d). Interestingly, a third noise peak at lower temperatures was also observed at $I_{ds} = 30$ µA and 50 µA. The lower temperature peak disappeared when measured at the higher current of $I_{ds} = 100$ µA, as shown in Figure 4 (d), and at $I_{ds} = 200$ µA (see also Supplemental Figure S5). The overall noise level remained low at higher temperatures as the device temperature approached RT.

When comparing the noise characteristics with the corresponding resistance data, one can conclude that the noise peak in the higher temperature region, *i.e.*, at $T\sim145$ K, corresponds to the onset of the resistive anomaly due to CDW phase transition where the material behavior changes due to a partial bandgap opening and removal of a part of the conduction electrons to the pinned quantum condensate[56,57,59]. The noise increase is associated with the appearance of the Lorentzian features (see Figure 3). The noise evolution from $1/f$ to the Lorentzian spectrum as a result of the CDW phase transition has been previously reported by us for quasi-1D and 2D materials[34,35,40]. The appearance of the Lorentzian at the phase transition points can be generally interpreted as due to the system, fluctuating between two phases at the transition point, before settling to one specific phase. The latter results in the Lorentzian, characteristic for the two-level systems where the material resistance switches between the higher and lower resistive states.

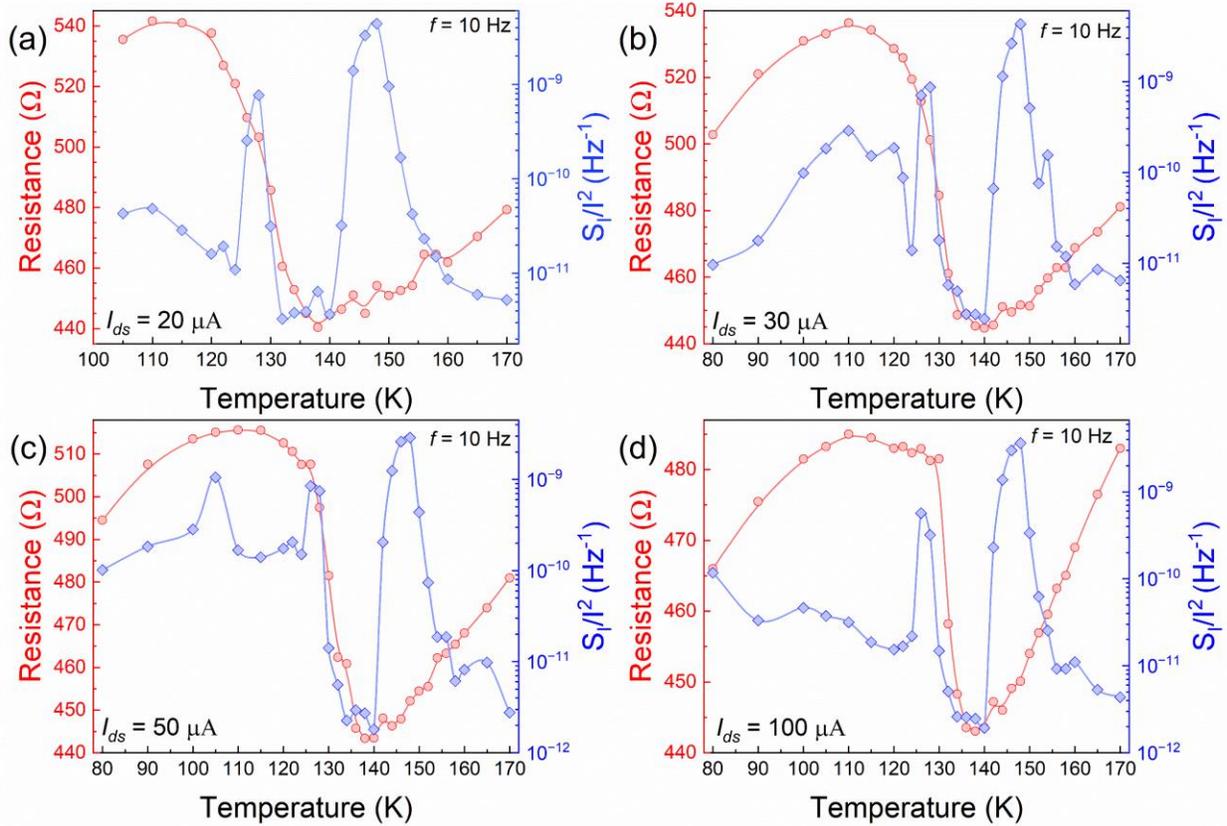

**Figure 4:** (a) Resistance (left axis) and the noise spectral density, $S_I/I^2$, (right axis) measured as the functions of temperature in the range that includes the CDW phase transition. The data are shown at a fixed frequency $f = 10$ Hz and the channel current $I_{ds} = 20$ µA. The noise vs. temperature data shows well-defined peaks at and below the CDW transition temperature. The panels (b), (c), and (d) are the same as panel (a) but show data for the different current levels $I_{ds} = 30$ µA, 50 µA, and 100 µA, respectively. Note the evolution of the noise spectral density and the corresponding changes in resistance.

The interpretation of two other peaks at lower temperatures is more complicated. Let us consider the noise peak at $T \sim 130$ K. The noise in the linear I-V regime is due to the individual particle conduction. One can assume that the initial increase in the noise with decreasing temperature is due to the reduction in the number of the conduction electrons since part of the electrons condensed into the pinned CDW. The 1/*f* noise of individual electrons scales inversely with the number of carriers, *i.e.*, $S_I/I^2 \sim 1/(N \times f)$, where $N$ is the total number of charge carriers in the conduction channel. However, it is difficult to explain the fast decrease of the noise at lower temperatures and the overall narrow shape of this second peak. One should also keep in mind that the increase of the noise in both peaks in Figure 4 (a-d) is associated with the emergence of the Lorentzian features in the spectra. When comparing this second noise peak with the corresponding resistance behavior, one can see that the temperature at which the second peak attains its highest value roughly corresponds to the temperature where the resistance change is maximum as a function of temperature, *i.e.*, *dR/dT* attains its maximum value (see also Supplemental Figure S6). The emergence of this second peak could be associated with the fast change in the resistance value of the device in the non-metallic phase below the phase transition. Thus, if the first peak signifies the material transition to the IC-CDW phase, the second peak appears around the temperature where the resistance changes are the strongest.

Another intriguing feature in the noise characteristics is the emergence of a third peak at lower temperatures, as observed in Figure 4 (b) and (c) at the current levels $I_{ds}$ = 30 µA and 50 µA, respectively. These current levels correspond to the bias voltages, which are close to the threshold field, $V_{th}$ (see Figure 2 (b)). At this *I-V* region, just before the CDW depinning, the carrier dynamics can already be affected by the CDW condensate motion, referred to as the CDW *creep* [68]. The term creep in the CDW context means an incoherent slow motion, with stops, of the CDWs or segments of CDWs, which precede the coherent CDW sliding. The CDW creep regime can be considered as a separate phase or sub-phase of the incommensurate CDW phase, which explains the appearance of Lorentzians. This explanation is consistent with the third peak disappearing at the higher current of $I_{ds}$ = 100 µA, away from the threshold, where the CDW is already sliding, and the noise is reduced. The presence of several peaks is an indication that the phase transition has a fine structure with the existence of several, at least one, intermediate phases, which can be seen only using noise spectroscopy analysis. These intermediate phases can be similar to the hidden phases observed in other CDW materials[69–71].

IV. **Electrical Field Dependence of Noise in NbSe$_3$ Nanowires**

We now turn to the electric-field dependence of noise in the NbSe$_3$ nanowire devices. Figures 5 (a), (b), and (c) show the *I-V* characteristics and the normalized noise spectral density, $S_I/I^2$, at a fixed frequency $f = 10$ Hz, in the IC-CDW phase at $T = 100$ K, 120 K, and 130 K, respectively. The normalized noise spectra density as a function of the bias voltage in Figure 4 (a) shows two peaks, the smaller one in the linear *I-V* regions below the threshold field, $V_{TH}$, and the large peak at $V_{TH}$ where *I-V*s become super-linear, and the CDW starts sliding. The smaller peak can be associated with the CDW creep, the slow, incoherent motion of CDW condensate before the onset of sliding[68] or intermediate hidden phases[70]. While it is not possible at the moment to distinguish between these two possible explanations of the smaller peak noise peak, we note that the CDW creep has been invoked to explain the characteristics of bulk NbSe$_3$ crystals in several prior reports[68,72,73]. The CDW creep occurs below the threshold field, $E_{TH}=V_{TH}/L$, in the linear *I-V* region when the applied field is sufficient to overcome some of the weaker pinning forces, leading to a partial CDW motion below $E_{TH}$. The creep can be described by its own critical electric field, $E_{TH}^*$ [68]. A similar behavior is observed in Figure 5 (b) and (c) at $T = 120$ K and 130 K. The separation between the two peaks is less apparent at $T = 130$ K because the measurement temperature, is close to $T_{P1}$, and the incommensurate CDW phase has just formed (see Figure 2 (c)). The larger noise peak at a slightly higher bias voltage corresponds to the depinning of the CDW and the onset of its sliding. Thus, the explanation of the two peaks observed in the $S_I/I^2$ vs. *V* dependence in Figure 5 is consistent with the explanation of the peaks in the $S_I/I^2$ vs *T* dependence in Figure 4. Additional noise data at other temperatures is presented in Supplemental Figure S7.

Figure 5 (d) shows the threshold voltage, $V_{TH}$, in the IC-CDW phase as a function of temperature. The value of $V_{TH}$ decreases with increasing temperature and reaches a minimum at $T\sim126$ K, after which $V_{TH}$ increases with temperature. The non-monotonic behavior of the threshold field with temperature has been previously observed for NbSe$_3$ [42,74] and other quasi-1D CDW materials[39]. Interestingly, the amplitude of the noise peak at the onset of depinning also shows a non-monotonic behavior, with its value initially increasing as the material temperature increases and dropping again at higher temperatures (see also Supplemental Figures S8 and S9). The behavior can be observed in Figure 5 (a-c), where the noise peak at $T=130$ K is lower than the values at $T = 100$ K and 120 K. One should note that multiple peaks in the temperature dependence and non-monotonic dependence of the threshold voltage on temperature can be interpreted as the support for the intermediate hidden phases in the temperature range where the resistance decreases with the temperature increase. The CDW creep is by itself is considered to be a metastable phase.

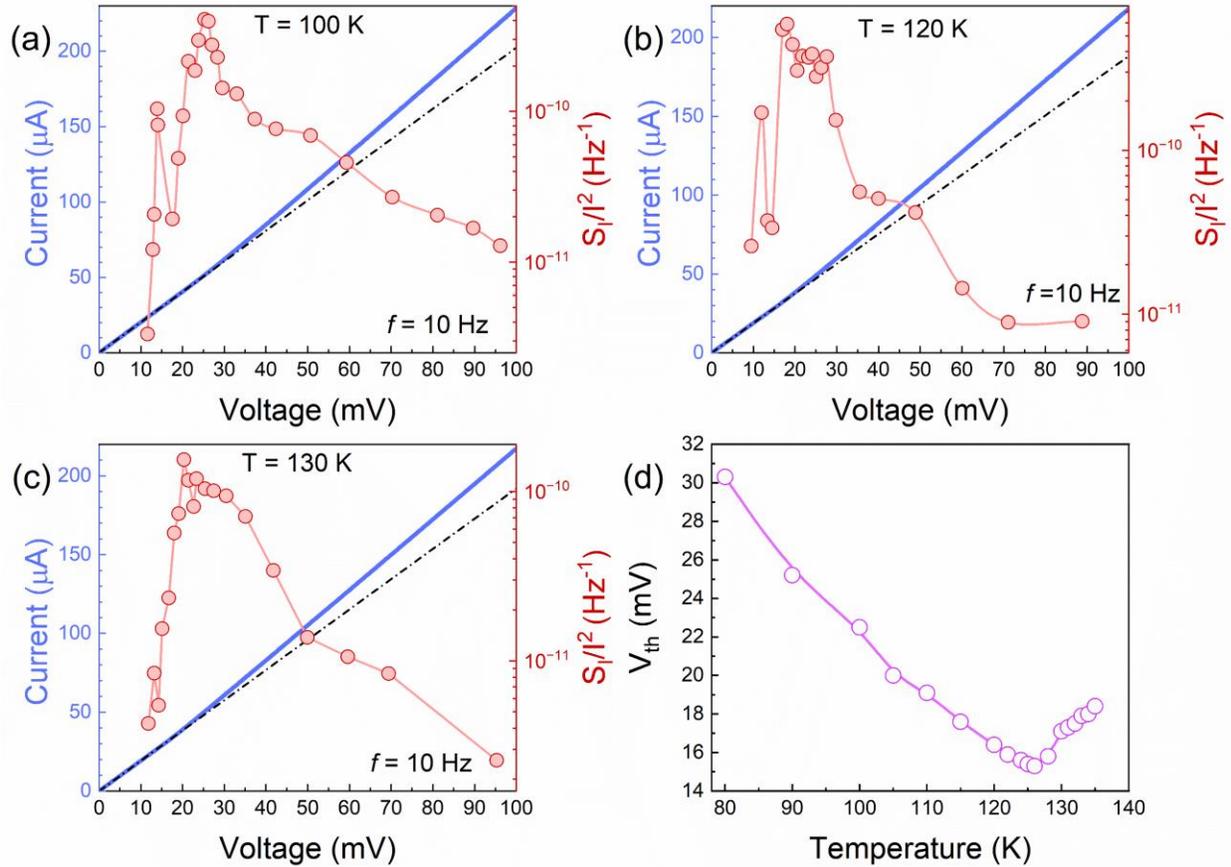

**Figure 5:** (a) Current-voltage (*I-V*) characteristics (left axis) and the normalized noise spectral density, $S_I/I^2$, at $f=10$ Hz (right axis), as a function of applied voltage. The data are shown for $T = 100$ K, below the CDW phase transition. The onset of the non-linear current corresponds to the depinning and sliding of the CDW quantum condensate. The noise spectra reveal two peaks – the smaller one slightly below the depinning field and the larger one at the onset of depinning ($I_{CDW} > 0$). The panels (b) and (c) are the same as panel (a) but show the data for temperatures $T=120$ K and 130 K, respectively. (d) The threshold voltage dependence on the temperature in the incommensurate CDW phase.

To further correlate the noise spectral features with the CDW carrier depinning and sliding, we analyzed the noise spectral density at different bias voltages. Figure 6 (a) shows the *I-V*s and $S_I/I^2$ ($f$ =10 Hz) as a function of bias voltage at $T$ =105 K (see also Supplemental Figures S10 and S11 for other temperatures). As discussed previously, the $S_I/I^2$ dependence on the bias voltage reveals two prominent noise peaks, one at the onset of the CDW depinning at $V_{TH}$ and another at the lower bias, in the linear region. The corresponding normalized noise spectral density *vs.* frequency plots are shown in Figures 6 (b), (c), and (d). In Figure 6 (b), the noise spectral density *vs.* frequency is shown in the linear *I-V* region, between 12 mV and 18 mV bias voltages, where the first low-bias peak appears, possibly due to CDW creep. The noise spectra contain pronounced Lorentzian bulges on the *1/f* envelope. Contrary to the Lorentzian components corresponding to the first peak, the characteristic frequency of the second peak Lorentzian is much lower, below the measurement limit. The Lorentzian component in the noise spectra increases the noise level and peaks across the voltage range. The noise spectra for the noise data points, corresponding to the second peak at the threshold field, are shown in Figure 6 (c). The noise spectra at the two bias voltages between 19 mV and 23 mV show similar Lorentzian bulges on the *1/f* background. With further increase in the bias voltage, the Lorentzian bulge disappears, as seen in the spectra measured under bias voltages 79.2 mV, and 96.6 mV, which are far from the CDW depinning point. The disappearance of the Lorentzian components at higher bias confirms that the appearance of bulges is related to the CDW carrier dynamics of NbSe$_3$ material and not external factors. Finally, in Figure 6 (d), we plotted $f \times S_I/I^2$ *vs. f* for different biases in the linear region to eliminate the 1/$f$ background and present the Lorentzian more clearly. One can see how the peaks move to higher frequencies with increasing bias voltage (see also Supplemental Figure S12). A similar trend was observed for a quasi-2D CDW material 1$T$-TaS$_2$ [75].

V. **The Noise of the Collective Current of the Sliding Condensate**

We now address the question of how noise in the collective current of the sliding quantum condensate compares to the noise of the current of the individual electrons. In Figure 7 (a-d), we show the noise contribution from the CDW condensate to separate it from the overall noise. The data are presented for two representative devices. Figure 7 (a) shows the *I-V* characteristics of the first device measured at 105 K. The total current, $I_T$, includes the current of individual, i.e., single, electrons, $I_S$, and the collective current of the sliding CDW condensate, $I_{CDW}$. The total noise spectral density is the sum of spectral densities of the noise of the individual electron current and the collective CDW current. In Figure 7 (b), we plotted the noise of each current type as a function of total current, $I_T$. At lower current levels, the spectral density of the total noise, $S_{IT}$, follows a conventional quadratic relationship with current. As the current increases, the total noise experiences an abrupt jump and reveals two peaks, as explained previously. After the second peak, the overall noise starts to drop. The normal carrier noise calculated from the noise behavior at the lower current levels follows a near quadratic behavior with the total current. The $S_{I_S}$ *vs.* $I_T$ shows a slight deviation from the quadratic behavior due to the emergence of CDW current. Most interestingly, the CDW noise, $S_{I_{CDW}}$, decreases with the increasing current after depinning, revealing a drastic drop at higher current levels. Intriguingly, the noise of the collective current goes even below the linear resistor noise. Therefore, the noise of the collective current of the sliding CDW condensate differs substantially from the noise of resistors and active electronic devices, *e.g.*, diodes and transistors[40,65–67]. A similar behavior of the noise of CDW current was observed in a different device measured intentionally at a slightly different temperature $T=100$ K (see Figure 7 (c) and (d)). One can conclude that the CDWs in the IC-CDW phase are noisy near the depinning bias voltage. As the sliding of the CDW becomes established, the noise decreases. It is important to note that at high currents, the noise of the normal electron current and total noise levels coincide. This indicates that the noise of the CDW collective current tends to disappear. The latter consideration creates an additional motivation for further investigation of the collective current in CDW materials for future applications in low-power, low-noise electronics.

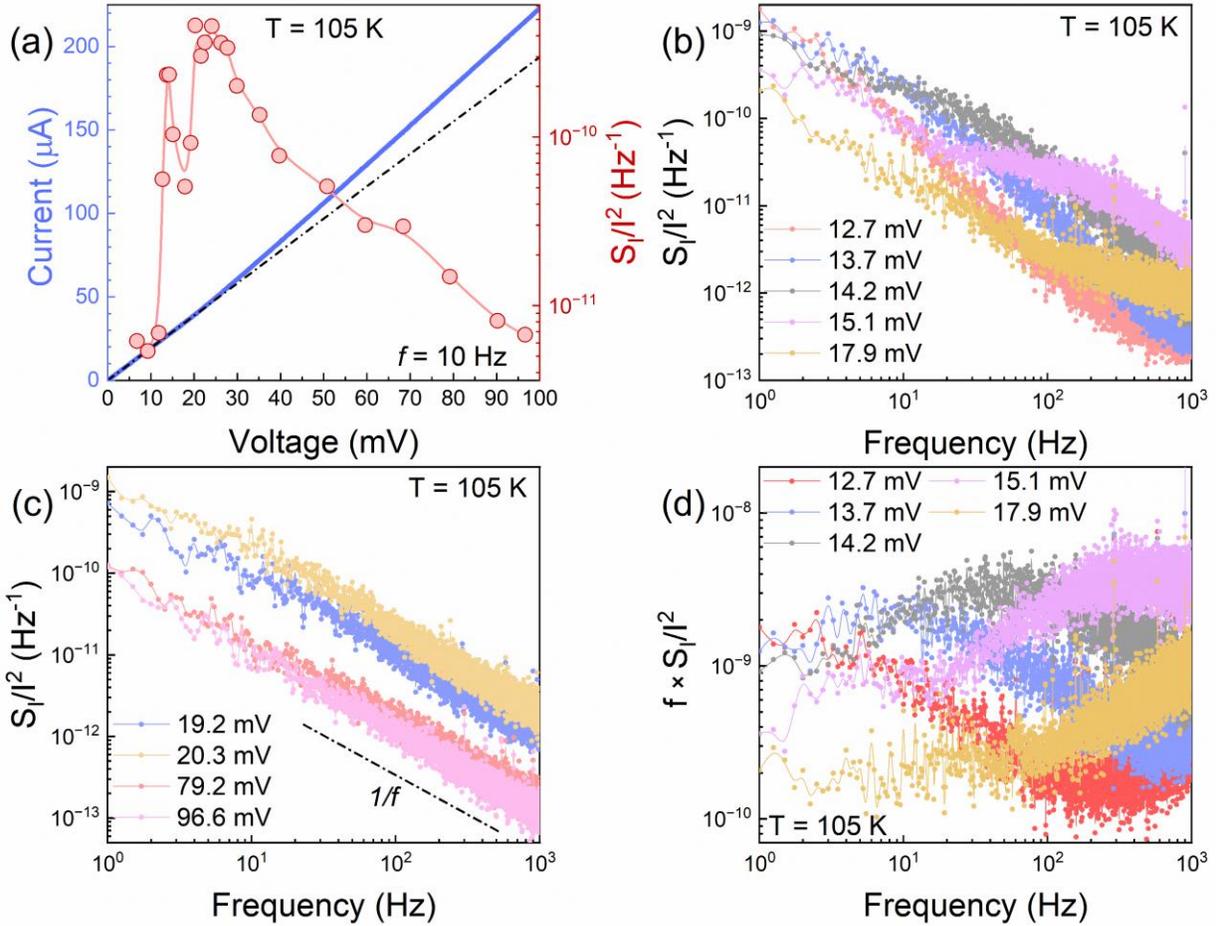

**Figure 6:** (a) Current (left axis) and the normalized noise spectral density, $S_I/I^2$, at $f=10$ Hz (right axis) as the functions of the applied voltage at the temperature $T=105$ K. Note the two peaks in the noise level. (b) The normalized noise spectral density, $S_I/I^2$, as a function of frequency at five different bias voltages, in the range where the noise level in panel (a) shows the first peak. The noise spectra reveal Lorentzian bulges at these bias voltages. (c) The $S_I/I^2$ vs. $f$ plots at four different bias points. The first two bias points correspond to the second noise peak in panel (a), where a new Lorentzian bulge appears in the noise spectra. The remaining two higher bias points correspond to the region where the noise spectrum became the $1/f$ type again. (d) The $f \times S_I/I^2$ vs. $f$ plots for the same bias voltage as panel (b). Note the evolution of the Lorentzian corner frequency, *i.e.*, shifting to the right, with the increasing bias voltage.

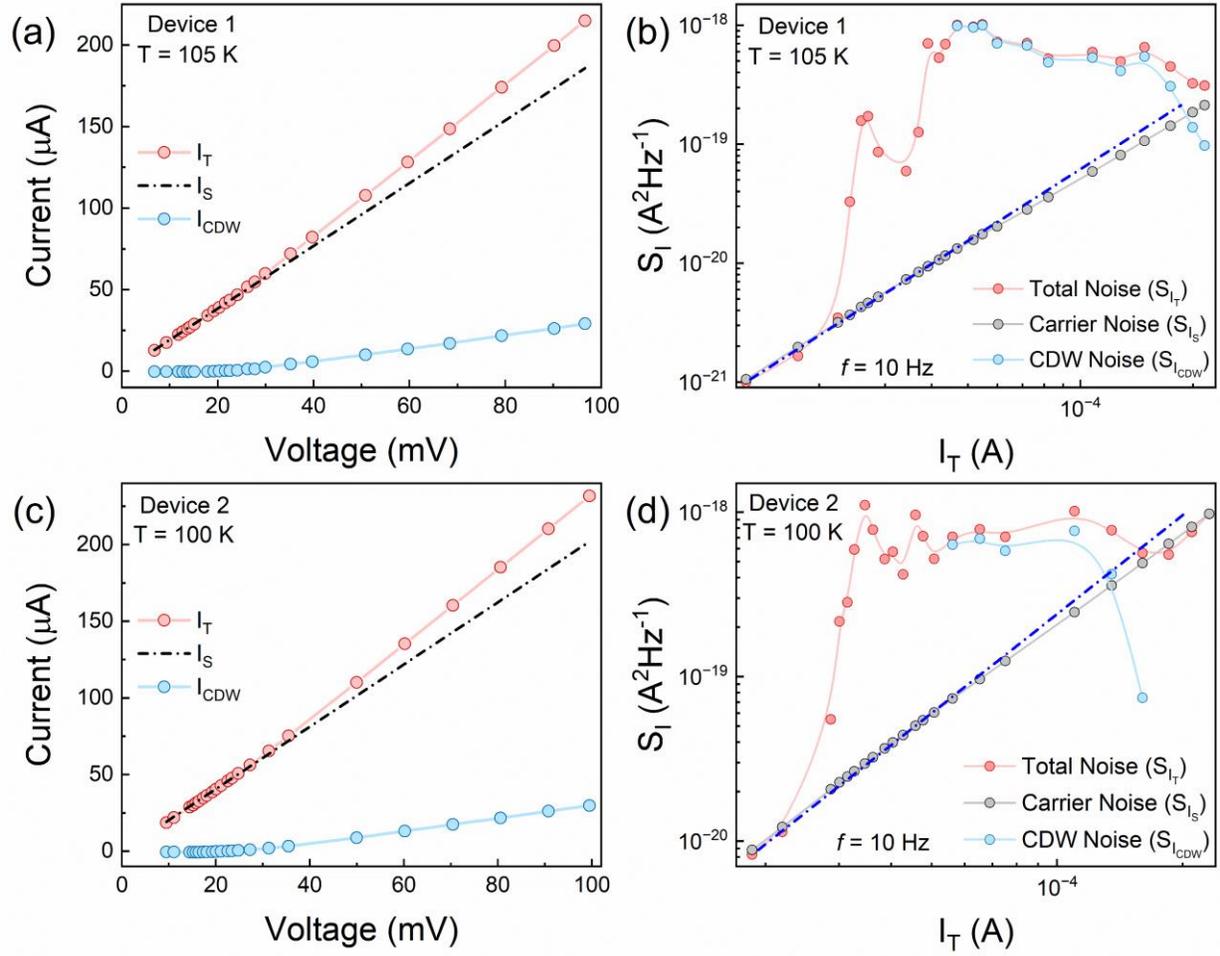

**Figure 7**: (a) Current-voltage (*I-V*) characteristics for representative device #1 at $T = 105$ K. The total current $I_T$ after depinning is equal to the combination of normal carrier current, $I_S$, and CDW current, $I_{CDW}$. (b) The current noise spectral density, $S_I$, at $f = 10$ Hz, *vs.* total current, $I_T$. Note that the noise of the sliding CDW condensate decreases with increasing current. The dashed line shows the quadratic relationship between the noise of the normal charge carriers and the current of normal carriers, *i.e.*, $S_{I_S} \sim I_S^2$. (c) The same as in panel (a) but for representative device #2 and slightly lower temperature $T = 100$ K. (d) The same as in panel (b) but plotted for representative device #2 and temperature $T = 100$ K. One can see again a drastic decrease in the noise of sliding quantum condensate with increasing current.

## VI. Conclusions

We investigated noise in nanowires of quasi-1D $NbSe_3$, focusing on a temperature range near the Pearls transition $T_{P1}$ ~145 K. The data analysis allowed us to separate the noise produced by the individual conduction electrons and the sliding CDW quantum condensate. We found that the noise of the collective current reduces with the increasing bias voltage in contrast to the constant noise of the current of the individual electrons. The obtained results shed light on the behavior of pinned and sliding charge density wave quantum condensate and demonstrate the potential of noise spectroscopy for investigating the properties of low-dimensional quantum materials. The reduction of the noise of CDWs with increasing current can be important proposed applications of CDW quantum materials in future low-power and radiation-hard electronics.

**Acknowledgments**

The work at UCLA was supported, in part, *via* the Vannevar Bush Faculty Fellowship (VBFF) to A.A.B. under the Office of Naval Research (ONR) contract N00014-21-1-2947 on One-Dimensional Quantum Materials. A.A.B. also acknowledges the support from the National Science Foundation (NSF) program Designing Materials to Revolutionize and Engineer our Future (DMREF) via a project DMR-1921958 entitled Collaborative Research: Data Driven Discovery of Synthesis Pathways and Distinguishing Electronic Phenomena of 1D van der Waals Bonded Solids. S.R. acknowledges partial support by the CENTERA Laboratories under the European Regional Development Fund No. MAB/2018/9 and by the European Union ERC "TERAPLASM" project No. 101053716. The authors thank Dr. Fariborz Kargar (UCLA) and Dr. Roger Lake (UCR) for their valuable discussions.

**Author Contributions**

A.A.B. conceived the idea, coordinated the project, contributed to experimental data analysis, and led the manuscript preparations. S.G. fabricated devices, measured *I-V*s, and noise characteristics, and analyzed the experimental data; S.R. contributed to experimental data analysis. All authors participated in the manuscript preparation.

**Supplemental Information**

The supplemental information is available at the Applied Physics Reviews journal website free of charge.

**The Data Availability Statement**

The data that support the findings of this study are available from the corresponding author upon reasonable request.